\documentclass{article}

\usepackage[usenames]{color}

\def\giorno{30/04/2014}

\def\a{\alpha}
\def\b{\beta}

\def\ga{\gamma}
\def\de{\delta}   
\def\eps{\varepsilon}

\def\s{\sigma}
\def\z{\zeta}
\def\om{\omega}

\def\C{{\bf C}}

\def\E{{\cal E}}

\def\G{{\cal G}}
\def\H{{\bf H}}

\def\h{{\cal H}}

\def\Jb{{\bf J}}

\def\L{{\cal L}}

\def\R{{\bf R}}

\def\T{{\rm T}}

\def\Y{{\mathcal Y}}

\def\La{\Lambda}
\def\Om{\Omega}

\def\E{\mathtt{I}}

\def\pa{\partial}

\def\d{{\rm d}}       
\def\w{\wedge}
\def\xb{{\bf x}}

\def\o+{\oplus}
\def\xd{{\dot x}}

\def\ss{\subset}

\def\<{\langle}
\def\>{\rangle}

\def\({\left(}
\def\){\right)}
\def\[{\left[}
\def\]{\right]}
\def\<{\langle}
\def\>{\rangle}

\def\=#1{\bar #1}
\def\~#1{\widetilde #1}
\def\wt#1{\widetilde #1}
\def\.#1{\dot #1}
\def\^#1{\widehat #1}
\def\"#1{\ddot #1}

\def\EOR{~ \hfill {$\odot$}}

\def\beq{\begin{equation}}
\def\eeq{\end{equation}}
\def\eqref#1{(\ref{#1})}

\def\interno{\hskip 2pt \vbox{\hbox{\vbox to .18
truecm{\vfill\hbox to .25 truecm
{\hfill\hfill}\vfill}\vrule}\hrule}\hskip 2 pt}

\def\hh{hyper\-hamil\-to\-nian }
\def\hk{hyper\-kahler }
\def\hs{hyper\-symp\-lec\-tic }

\def\torus{{\bf T}}

\begin{document}

\title{Symmetry and quaternionic integrable systems}

\author{G Gaeta\thanks{Dipartimento di Matematica, Universit\`a degli Studi di Milano, via Saldini 50, I-20133 Milano (Italy); e-mail: {\tt giuseppe.gaeta@unimi.it}} \ and M A Rodr\'{\i}guez\thanks{Departamento de F\'{\i}sica Te\'orica II, Universidad Complutense, E-28040 Madrid (Spain); e-mail: {\tt rodrigue@fis.ucm.es}}}

\date{\giorno}

\maketitle

\noindent
{\bf Abstract.} Given a hyperkahler manifold $M$, the hyperkahler structure defines a triple of symplectic structures on $M$; with these, a triple of Hamiltonians defines a so called hyperhamiltonian dynamical system on $M$. These systems are integrable when can be mapped to a system of quaternionic oscillators. We discuss the symmetry of integrable hyperhamiltonian systems, i.e. quaternionic oscillators; and conversely how these symmetries characterize, at least in the Euclidean case, integrable hyperhamiltonian systems.

\section*{Introduction}

Integrable finite dimensional Hamiltonian systems are widely
studied, but the Hamiltonian class does not exhaust the set of
integrable systems. Here we are concerned with a class of
integrable systems which generalizes Hamiltonian ones, in a sense
to be detailed below, i.e. with {\it quaternionic integrable
systems}.

It is well known that Hamiltonian integrable systems are
characterized by a high degree of symmetry \cite{Arn1,Gall}; here
we will investigate the symmetries of an integrable quaternionic
system.

We will be interested in systems (both quaternionic and, in
drawing parallels and detailing differences with these,
Hamiltonian) such that the accessible set of spaces for finite
initial conditions is a compact manifold; for Hamiltonian systems
this means that energy manifolds are compact (and this is the
``most interesting'' case, at least from the point of view of
\cite{EMS3}; see chapter 5.1 (page 175) there). The same will
hold for quaternionic systems, with a suitable definition of Energy.

It should be mentioned that quaternionic integrable systems arise
as a special class of {\it hyperhamiltonian dynamical systems};
the latter represent a generalization of Hamiltonian ones, with
the role of the symplectic structure taken by a hypersymplectic
structure. This in turn is the symplectic counterpart to a
hyperkahler structure (i.e. a certain set of Kahler structures,
see below). This means that one should have in mind the case of
Hamiltonian systems with a Kahler structure rather than with a
generic symplectic one; in other words, we will explore a
generalization of Hamiltonian systems for which a metric exists
and is preserved.

Finally, albeit the present work will be at a purely mathematical level, we
devote some word to the physical motivation behind the (introduction and)
study of hyperhamiltonian dynamics. This is primarily provided by systems with spin; in
particular, in previous work we have shown that the Pauli and the Dirac
equations can be cast in hyperhamiltonian form \cite{GR}; it is also known
that the Pauli equation corresponds to an integrable hyperhamiltonian
system \cite{Gdeg}, while integration of the hyperhamiltonian flow
corresponding to the Dirac equation requires consideration of dual
hyperkahler structures (see below in this paper).

Separation of the Dirac equation into two equations (one for the positive and one for the
negative energy states) up to some order in a perturbation expansion can
also be recast in hyperhamiltonian formalism, both in the Foldy-Wouthuysen
approach \cite{BD,FW} (non-relativistic limit) and in the Cini-Touschek one \cite{CT}
(ultra-relativistic limit; see also Mulligan \cite{Mul}), is also possible within the
hyperhamiltonian formalism \cite{GR}.

Hyperkahler systems are also relevant in different physical contexts, in particular
in General Relativity \cite{EGH}; e.g. it is known that Taub-NUT spaces are hyperkahler
(and can be obtained from $\R^8$ with standard quaternionic structure, see below, through the
HKLR quotient procedure \cite{HKLR}). We mention in this regard that the complex structures
for Taub-NUT spaces, which are a required input to formulate hyperhamiltonian dynamics on these, have been
recently computed in fully explicit form \cite{GRtnut}.

\bigskip

The {\it plan of the paper} is as follows. In section \ref{sec:back}
we will briefly recall some well known notions in order to fix notation;
in particular, we recall some basic features of Hamiltonian integrable
systems in subsection \ref{sec:hamint}, the notion of Kahler manifold in
subsection \ref{sec:kahler}, and that of hyperkahler manifold
in section \ref{sec:hkman}.
In section \ref{sec:hh} we will introduce \hh dynamics; and in section \ref{sec:hhint} we will
define and discuss \hh (quaternionic) integrable systems, together with a generalization
of these, i.e. Dirac oscillators (see section \ref{sec:dirac}). At this
point we will have all the needed background, and be ready to
start discussing our problem. In particular, in section
\ref{sec:symmint} we will analyze the symmetries of a \hh
integrable systems, while in section \ref{sec:intsymm} we will
discuss if symmetries do characterize integrable systems in the
\hh case. In the final section \ref{sec:concl} we will summarize
our findings and draw conclusions as well as discuss relations
with other works in related subjects.
\bigskip

As for {\it notation}, in this paper we will use Greek indices $\a,\b,...$ (taking values 1,2,3) associated to the
\hk structure on $4n$-dimensional manifolds, and Latin indices (taking values $1,...,4n$)
associated to the spatial local coordinates on the manifolds; summation with respect to
repeated (upper and lower) Latin indices will always be understood, while summation with
respect to Greek indices will be explicitly indicated to avoid confusion.

\bigskip\noindent {\bf Acknowledgements.} GG is supported by the Italian MIUR-PRIN
program under project 2010-JJ4KPA. MAR is supported by the Spanish Ministry of Science and Innovation under
project FIS2011-22566.

\section{Background}
\label{sec:back}

In this section we collect well known notions and constructions, to be used in the following; this will also serve to set our notation, to be freely used below.

\subsection{Integrable Hamiltonian systems}
\label{sec:hamint}

Finite-dimensional integrable Hamiltonian systems are widely studied and well known; thus here we will just recall some basic features, mainly to fix notation but also to illustrate the point of view which leads more directly to the generalization considered in this paper, i.e. integrable \hh systems.

Here we understand integrability in Arnold-Liouville sense; thus a system in $n$ degrees of freedom -- equivalent to a system of $2 n$ first order ODEs -- is integrable if it can be mapped (via a diffeomorphism) to an $n$-dimensional harmonic oscillator \cite{Arn1,Gall},
\beq \dot{q}_k \ = \ \nu_k \ p_k \ , \ \ \dot{p}_k \ = \ - \, \nu_k \, q_k \ ; \eeq
here the $\nu_k$ only depend (possibly) on the quantities $I_k = (p_k^2 + q_k^2)$. Note that there is no sum on repeated indices, and the same will apply in all of this section.

If we pass to action-angle coordinates $(I,\phi)$ via
\beq p_k \ = \ \sqrt{I_k} \ \cos{\phi_k} \ , \ \ q_k \ = \ \sqrt{I_k} \ \sin{\phi_k}  \eeq (this transformation is singular in the origin) then the evolution reads
\beq \dot{I}_k \ = \ 0 \ , \ \ \dot{\phi}_k \ = \ \nu_k \ . \eeq
Equivalently, we can use complex coordinates
\beq z_k \ = \ p_k \ + \ i \, q_k \ \equiv \ \sqrt{I_k} \ \exp [ i \, \phi_k] \ ; \eeq now the evolution reads
\beq \dot{z}_k \ = \ i \ \nu_k \ z_k \ . \eeq
The frequencies $\nu_k$ are in general a function of the $I_k$, equivalently of the $|z_k|^2$, variables.

Two features are immediately apparent:
\begin{itemize}
\item[(1)] The system is invariant under the abelian group \beq \mathrm{SO(2)} \times ... \times \mathrm{SO(2)} \ \equiv \ \mathrm{U(1)} \times ... \times \mathrm{U(1)} \ = \ \torus^n \ ; \eeq
\item[(2)] time evolution is given by a (real or complex) rotation, with speed $\nu_k$, in each ($\R^2$ or $\C^1$) subspace.
\end{itemize}

It is well known that, conversely, a system enjoying a $\torus^n$ symmetry is integrable (in Arnold-Liouville sense) \cite{Arn1}, and it is obvious that if the phase space can be fibred in terms of two-dimensional manifolds $M^2$ as $M^2 \times ... \times M^2$, or in terms of complex lines $\C^1$ as $\C^1 \times ... \times \C^1$, with time evolution described by rotations in each factor, then the system is integrable.

The heuristic idea behind quaternionic integrable systems will be to replace complex rotations (in $\C^1 \simeq \R^2$ spaces) with quaternionic rotations (in $\H^1 \simeq \R^4$ spaces).

\subsection{Kahler manifolds}
\label{sec:kahler}

As well known, a symplectic manifold is not required to carry any metric.
On the other hand, let us consider a smooth, $2n$-dimensional, real
manifold $M$ equipped with a Riemannian metric $g$. This also defines a
canonical connection on it, i.e. the unique torsion-free Levi-Civita connection
associated to the metric $g$; we will denote it by $\nabla$.

An almost complex structure on $(M,g)$ is a (1,1) type tensor
field $J$ such that $J^2 = - I$, with $I$ the identity map.

A {\it Kahler manifold } $(M,g,J)$ is a smooth orientable real
Riemannian manifold $(M,g)$ of dimension $m=2n$ equipped with an
almost-complex structure $J$ which has vanishing covariant
derivative under the Levi-Civita connection, $\nabla J = 0$.
The latter condition implies (as stated by the
Newlander-Nirenberg theorem \cite{NN}) the integrability of
$J$; so $(M,g,J)$ is a complex manifold.

The two-form $\om \in \La^2 (M)$ associated to $J$ and $g$ via the
Kahler relation \beq\label{eq:kahler} \om (v,w) \ = \ g (v,Jw)
\eeq is a symplectic form; hence each Kahler manifold is also symplectic.
(The converse is not true, and there are symplectic manifolds
which do not admit any Kahler structure.)

\subsection{Hyperkahler manifolds}
\label{sec:hkman}

A {\it hyperkahler manifold } is a real smooth orientable
Riemannian manifold $(M,g)$ of dimension $m = 4n$ equipped with
three complex structures\footnote{Thus, in particular, these will satisfy
$\nabla J_\a = 0$.} $J_1,J_2,J_3$ which satisfy the
quaternionic relations
\beq\label{eq:quaternionic} J_\a \,
J_\b \ = \ \sum_{\gamma}\epsilon_{\a \b \ga} \, J_\ga \ - \ \delta_{\a \b} \, I \ ;
\eeq here $\epsilon_{\a \b \ga}$ is the completely antisymmetric
(Levi-Civita) tensor, and $\de_{\a \b}$ the Kronecker symbol.
The ordered triple ${\Jb} = (J_1 , J_2 , J_3)$ is a {\it
hyperkahler structure } on $(M,g)$.

Simple examples of hyperkahler manifolds are provided by
quaternionic vector spaces ${\bf H}^k$ and by the cotangent bundle
of complex manifolds.

Note that the quaternionic relations imply that the $J_\a$ satisfy the $\mathrm{SU(2)}$
commutation relations, but also involve the multiplication structure.

Obviously a hyperkahler manifold is also Kahler with respect to
any linear combination $J = \sum_\a c_\a J_\a$ of the $J_\a$ with $|c|^2 :=
c_1^2 + c_2^2 + c_3^2 = 1$. Moreover, as the Kahler structures identify
symplectic ones, to the triple $(J_1,J_2,J_3)$ corresponds a
triple $(\om_1 , \om_2 , \om_3)$ of symplectic structures via
$\om_\a (v,w) := g (v,J_\a w)$; we will then speak of a \hs structure.
Actually, any linear combination  $\om = \sum_\a c_\a \om_\a$ of the $\om_\a$ with $|c|^2 :=
c_1^2 + c_2^2 + c_3^2 = 1$ will be a symplectic form on $M$.

\medskip\noindent
{\bf Remark 1.}
The space \beq\label{eq:quatstruct} {\bf Q} \ := \ \big\{ \sum_\a c_\a J_\a
\ , \ c_\a \in {\bf R} \big\} \ \approx \ {\bf R}^3 \ , \eeq will be
called the {\bf quaternionic structure} on $(M,g)$ spanned by
$(J_1,J_2,J_3)$ \cite{AlM}; two hyperkahler structures
on $(M,g)$ defining the same quaternionic structure ${\bf Q}$
will be seen as equivalent. An equivalence class of hyperkahler
structures is identified with the corresponding quaternionic
structure, and viceversa. The maps preserving the quaternionic structure
(i.e. carrying a given \hk structure into an equivalent one) are considered
as the {\it canonical maps} for the quaternionic structure; see \cite{GRcanonical}
for their characterization, and \cite{GReuclidean} for a fully explicit discussion
in Euclidean $\R^{4n}$ spaces. Note canonical maps induce necessarily a map
$$ J_\a \ \to \ \wt{J}_\a \ = \ \sum_\b \, R_{\a \b} \, J_\b $$ on complex structures,
with $R$ a real matrix in $\mathrm{SO(3)}$. \EOR

\subsection{The coordinate picture; standard structures in $\R^4$.}
\label{sec:standard}

It may be worth providing a description in local coordinates, also to fix notation to be widely used in the following. The complex structures $J_\a$ are represented by $(1,1)$ type tensor fields $Y_\a$; that is, with local coordinates $x^i $ ($i = 1 , ... , 4n$) we have\footnote{As anticipated in the  Introduction, sum over equal upper and lower Latin indices is understood from now on.}
$ J_\a = (Y_\a)^i_{\ j} \pa_{x^i} \otimes \d x^j$. The symplectic forms $\om_\a$ are represented by $(0,2)$ type antisymmetric tensor fields; that is, $\om_\a = (1/2) (K_\a)_{ij} \d x^i \w \d x^j$. The Kahler relation implies that $ K_\a \= g  Y_\a$;  in terms of the $K_\a$, the quaternionic relations are
$K_\a g^{-1} K_\b = \sum_{\gamma}\eps_{\a \b \ga}   K_\ga - \de_{\a \b} g$.

In the following it will also be convenient to consider $(2,0)$ type tensor fields $M_\a$ associated to the $Y_\a$ via $M_\a = Y_\a g^{-1}$; in terms of these the quaternionic relations stipulate
$ M_\a g M_\b = \sum_{\gamma} \eps_{\a \b \ga}  M_\ga - \de_{\a \b} g^{-1}$.

Note that the matrices $K_\a$, being associated to components of a differential two-forms, are antisymmetric; the same is immediately seen to hold in general for the $M_\a = g^{-1} K_\a g^{-1}$. Note also that in the Euclidean case $g = I$, and $M_\a = Y_\a = K_\a$; this means in particular that in this case the $Y_\a$ will be antisymmetric matrices.

\medskip\noindent
{\bf Remark 2.} The simplest example of hyperkahler manifold is that of $\R^4$ with euclidean metric. In this there are two standard hyperkahler structures, corresponding to the standard real representations of the Lie algebra $\mathrm{su(2)}$ and differing for their orientation. In fact, the real version of Schur Lemma (see e.g. \cite{Kir}, chap.8) states that real irreducible group representations are of three types: real, complex and quaternionic; for the latter case -- of interest here -- there exist two equivalent (and oppositely oriented) mutually commuting real representations. We will consider these as dual to each other, and correspondingly we will have a concept of dual \hk structures; note these share the same Riemannian metric. \EOR
\bigskip

In terms of the standard global coordinates $\{ x^1 , x^2,x^3,x^4 \}$ in $\R^4$, the complex structures of the positively-oriented standard \hk structure are given by
{\small \beq\label{eq:standhKpos} \Y_1 =
\pmatrix{0&1&0&0\cr -1&0&0&0\cr 0&0&0&1\cr 0&0&-1&0\cr}, \quad \Y_2
= \pmatrix{0&0&0&1\cr 0&0&1&0\cr 0&-1&0&0\cr -1&0&0&0\cr},\quad
\Y_3 = \pmatrix{0&0&1&0\cr 0&0&0&-1\cr -1&0&0&0\cr 0&1&0&0\cr} \ .
\eeq } The corresponding symplectic structures are given by
\beq\label{eq:standhSpos}
\begin{array}{ll}
\om_1 \ = \ \d x^1 \w \d x^2 \, + \, \d x^3 \w \d x^4 \ , &
\om_2 \ = \ \d x^1 \w \d x^4 \, + \, \d x^2 \w \d x^3 \ , \\
\om_3 \ = \ \d x^1 \w \d x^3 \, + \, \d x^4 \w \d x^2 \ . &
\end{array} \eeq

The negatively-oriented standard \hk structure is given by
{\small \beq\label{eq:standhKneg}
\widehat\Y_1 = \pmatrix{0&0&1&0\cr 0&0&0&1\cr
-1&0&0&0\cr 0&-1&0&0\cr}, \quad \widehat\Y_2 = \pmatrix{0&0&0&-1\cr
0&0&1&0\cr 0&-1&0&0\cr 1&0&0&0\cr}, \quad \widehat\Y_3 =
\pmatrix{0&-1&0&0\cr 1&0&0&0\cr 0&0&0&1\cr 0&0&-1&0\cr } \eeq }
To these correspond the symplectic structures
\beq\label{eq:standhSneg}
 \begin{array}{ll}
\^\om_1 \ = \ \d x^1 \w \d x^3 + \d x^2 \w \d x^4 \ , &
\^\om_2 \ = \ \d x^4 \w \d x^1 + \d x^2 \w \d x^3 \ , \\
\^\om_3 \ = \ \d x^2 \w \d x^1 + \d x^3 \w \d x^4 \ . &
\end{array} \eeq

With $\Om = \d x^1 \w \d x^2 \w \d x^3 \w \d x^4$ the standard volume form in $\R^4$, we note that the symplectic structures introduced above satisfy (no sum on $\a$)
\beq \label{eq:posor} (1/2) \ \( \om_\a \w \om_\a \) \ = \ \Om \eeq
for the positively-oriented ones, while for the negatively-oriented ones we have\footnote{These equations explain the notion of ``positively oriented'' or ``negatively oriented'' \hk structure.}
\beq \label{eq:negor} (1/2) \ \( \^\om_\a \w \^\om_\a \) \ = \ - \, \Om \ . \eeq
Note also that (in agreement with real Schur Lemma, see above) we have, for all $\a$ and $\b$,
$[\Y_\a , \widehat\Y_\b ] = 0$.

In the following, we will routinely use the notation $$ \pa_i \ \equiv \pa_{x^i}\ \equiv  \ (\pa / \pa x^i ) \ . $$

\medskip\noindent
{\bf Remark 3.} It is simple to see that given any set of matrices $L_\a$ satisfying the quaternionic relations \eqref{eq:quaternionic}, the associated symplectic forms necessarily define the same orientation, see \eqref{eq:posor} and \eqref{eq:negor}. These $L_\a$ can always be reduced via an orientation preserving orthogonal linear transformation in $\R^4$ -- i.e. a map $\wt{x}^i = R^i_{\ j} x^j$ with $R \in \mathrm{SO(4)}$ -- to either the set $\{ \Y_\a \}$ or the set $\{ \widehat\Y_\a \}$, depending on the orientation. \EOR

\section{Hyperhamiltonian systems}
\label{sec:hh}

In \hh dynamics one considers a $4n$-dimensional real manifold $M$ equipped with a Riemannian metric $g$ and three almost-integrable complex structures $J_\a$ making up a {\it \hk structure}. To this is associated, as mentioned above, a \hs structure; i.e. three symplectic structures $\om_\a$.

A \hh system is defined by an ordered triple of Hamiltonians $\h_\a$; each of this defines a Hamiltonian vector field $X_\a$ via pairing with the corresponding symplectic structure $\om_\a$. That is, we have the three vector field $X_\a$ satisfying
\beq X_\a \interno \om_\a \ = \ \d \, \h_\a \ . \eeq
The \hh vector field corresponding to the triple $(\h_1,\h_2,\h_3 )$ is
\beq \label{eq:hhvf} X \ = \ \sum_{\a = 1}^3 \, X_\a \ . \eeq

In local coordinates, freely using the notation introduced in the previous section, we have
$$ X_\a \ = \ f_\a^i \, \pa_i \ = \ (M_\a)^{ij} \, (\pa_j \h_\a) \ \pa_i \ ; $$
and therefore
$$ X \ = \ f^i \, \pa_i \ = \ \[ \sum_\a \, (M_\a)^{ij} \, (\pa_j \h_\a) \] \ \pa_i \ . $$
In other words, the equations of motion will be
\beq\label{eq:motion} \dot{x}^i \ = \ \sum_\a \, (M_\a)^{ij} \, (\pa_j \h_\a)  \ . \eeq

\medskip\noindent
{\bf Remark 4.} Hyperhamiltonian dynamics was introduced in \cite{GM}, actually motivated precisely by the integrable case \cite{GMspt} to be studied here, as a generalization of Hamiltonian dynamics in geometrical terms (that is, passing from a symplectic structure to a \hs one). It was shown in \cite{MT} that this formulation is also natural from the point of view of generalizing the complex structure formulation of Hamiltonian dynamics (in Kahler manifolds). Obviously, Hamiltonian systems are a special case of \hh ones (with two of the three hamiltonians $\h_\a$ being zero); it is easy to show -- e.g. by explicit example -- that there are \hh vector fields which are not Hamiltonian with respect to any symplectic structure \cite{GM} (this is based on necessary conditions for a vector field to be Hamiltonian with respect to some unspecified symplectic structure identified by Giordano, Marmo and Rubano \cite{GMR}). It was shown that \hh has a variational structure, albeit the variational principle leading to it is a non-standard one \cite{GM2}. Physically relevant equations -- in particular, as quite natural, those for particles with spin -- can be given a \hh structure; this is the case for the Pauli equation \cite{GM,GMspt,GR} and also for the Dirac equation \cite{GR}. In the latter case, actually, both \hk structures of a dual pair -- in this context, the two correspond to opposite helicity states -- enter in the description; the discussion makes use of a factorization principle for such a  dynamics (see Remark 5 below), based on previous work by Walcher \cite{Wal}. More recently, the \hk structure for Taub-NUT space \cite{NUT,Taub} was identified\footnote{Surprisingly, the literature only provided a discussion of the Taub-NUT metric, which is known on the basis of general argument to support a \hk structure (see also \cite{HKLR}), but the explicit form of the three complex structures seems not to have appeared previously to \cite{GRtnut}.} \cite{GRtnut}, which allows for studying \hh dynamics in this context, which is the simplest non-euclidean setting. In recent work, the theme of canonical maps for \hk structures -- and for quaternionic ones -- was also tackled; the general structure of the group of canonical transformations (which cannot be defined via a naive generalization of those for the symplectic case) has been determined\footnote{The results obtained in this context reproduce results which were already known via geometric constructions \cite{Be55,Joy}, but the discussion in \cite{GRcanonical} is conducted at an elementary level and reduces the problem to one in representation theory.} \cite{GRcanonical}, and a completely explicit description provided in the Euclidean case \cite{GReuclidean}. \EOR

\medskip\noindent
{\bf Remark 5.} In some contexts, in particular when dealing with the Dirac equation (see the previous Remark 4) one is led to consider vector fields in $M$ of the form
$$ X \ = \ X^{(+)} \ + \ X^{(-)} \ , $$
where $X^{(\pm)}$ are \hh with respect to dual \hk structures on $(M,g)$. As discussed in detail in \cite{GR}, it follows from the commutation of dual $\mathrm{SU(2)}$ representations (see Remark 2) that $$ [X^{(+)} , X^{(-)}] \ = \ 0 \ . $$ This, in turn, allows to use the factorization principle due to Walcher \cite{Wal}. Denoting by $\Phi(t;x_0;Y)$ the time $t$ flow issuing from $x_0$ at time $t=0$ under the vector field $Y$, we have
$$ \Phi(t;x_0;X) \ = \ \Phi\[t; \Phi(t;x_0;X^{(+)} ); X^{(-)} \] \ = \ \Phi\[t; \Phi(t;x_0;X^{(-)} ); X^{(+)} \] \ . $$
This means in particular that if we can integrate the flow under $X^{(\pm )}$, we also integrate immediately the flow under $X = X^{(+)} + X^{(-)}$. \EOR

\medskip\noindent
{\bf Remark 6.} Any \hh system in $\R^{4 n}$ with Hamiltonians $\{ \h_1 , \h_2 , \h_3 \}$ has a conserved $(4n - 1)$-form $\Theta$; denoting by $\z_\a$ the $(4n-2)$-forms $\zeta_\a = \om_\a \wedge ... \wedge \om_\a$ (with $2n-1$ factors), this is defined by
$$ \Theta \ = \ \sum_\a \d \h_\a \wedge \zeta_\a \ . $$
In $\R^{4 n}$ there is a natural correspondence between vector fields and $(4n-1)$ forms (through Hodge duality). Given $\chi \in \La^{4n-1} (\R^{4n})$, we will denote by $Y = F ( \chi)$ the corresponding vector field; this satisfies $Y \interno \Omega = \chi$. Given two forms $\chi , \eta \in \La^{4n-1} (\R^{4n})$, these define vector fields $Y_\chi = F (\chi)$ and $Y_\eta = F(\eta)$; the commutator $Z = [Y_\chi , Y_\eta]$ of these vector field is associated to a form $\psi \in \La^{4n-1} (\R^{4n})$, $\psi = F^{-1} (Z)$. Through this construction one defines a bracket $\{ . , . \} : \La^{4n-1} (\R^{4n}) \times \La^{4n-1} (\R^{4n}) \to \La^{4n-1} (\R^{4n})$; if $\chi$ and $\eta$ are conserved under the \hh dynamics, then $\psi = \{ \chi , \eta \}$ is also conserved \cite{GM}. The canonical form $\Theta$ also allows to characterize the \hh dynamics through a (non-standard) variational principle; see \cite{GM,GM2,GMdga} for details. \EOR

\section{Integrable hyperhamiltonian systems}
\label{sec:hhint}

A \hh system will be said to be {\it integrable} if it can be mapped to a system of {\it quaternionic oscillators} \cite{GMspt,Gdeg}.

We have mentioned above (see Remark 4) that there are \hh systems which are not Hamiltonian with respect to any symplectic structure; one may still wonder if there are {\it integrable} \hh systems which are not Hamiltonian. The answer to this question is affirmative, as was shown by explicit example in \cite{Gdeg}. Thus, it makes sense to investigate \hh systems.

\subsection{Quaternionic oscillators}
\label{sec:introQO}

A simple quaternionic oscillator in the Euclidean $\R^4$ space\footnote{All the $\R^{4n}$ spaces to be met in the following will be Euclidean; we will thus omit to specify this each time for ease of discussion; the $4n$-dimensional identity matrix will be denoted by $\E_{4n}$ (no confusion should be possible with the action variables $I_1,I_2,I_3$).} is a four-dimensional system of first-order ODEs of the form
\beq\label{eq:simplequatosc} \xd^i \ = \ \sum_{\a=1}^3  \, c_\a (|\xb|^2) \, (L_\a)^i_{\ j} \ x^j \eeq
with the real matrices $L_\a$ satisfying the quaternionic relations \eqref{eq:quaternionic}. The quantity
$$ \nu (|\xb|^2) \ = \ \sqrt{ c_1^2 (|\xb|^2) \ + \ c_2^2 (|\xb|^2) \ + \ c_3^2 (|\xb|^2) } $$ is the {\it frequency} of the oscillator.\footnote{If the $c_\a$ do not actually depend on $|\xb|^2$, i.e. are constant, we can always reduce to a Hamiltonian system. Note that one could have a constant $\nu$ even with non-constant $c_\a$.}

When working in $(\R^{4n},I)$ with coordinates $(x^1 , ... , x^{4n})$, it is convenient to introduce four-dimensional vectors
$\{ \xi_{(1)} , ... \xi_{(n)} \}$, with components
$$ \xi_{(k)}^i \ = \ x^{4(k-1) + i} \ . $$

A general quaternionic oscillator with $n$ degrees of freedom is a $4n$-dimensional system of first-order ODEs of the form
\begin{eqnarray} \label{eq:quatosc} \dot\xi_{(k)}^i &=& \sum_{\a=1}^3   \, c_\a (|\xi_1|^2, ... , |\xi_n|^2) \ (Y^{(k)}_\a)^i_{\ j} \ \xi_{(k)}^j \\ & & (i,j=1,...,4 ; \ k = 1,...,n ) \ , \nonumber \end{eqnarray} with the real matrices $Y_\a^{(k)}$ satisfying the quaternionic relations \eqref{eq:quaternionic}. In other words, we require to have an array of simple quaternionic oscillators, one in each $\R^4$ subspace, interacting only through their frequencies.

\medskip\noindent
{\bf Remark 7.} Needless to say, we can characterize quaternionic oscillators also without resorting to the $\xi_{(i)}$ vectors constructions; we then have general quaternionic oscillators in the form
\beq\label{eq:quatoscbis} \dot{x}^i \ = \ \sum_\a  \, c_\a [(\xb \cdot B_1 \xb) , ... , (\xb \cdot B_n \xb)] \ (L_\a)^i_{\ j} \ x^j \ .  \eeq Now $B_k$ is a sparse matrix having as only nonzero elements those on the diagonal at positions from $(4(k-1) + 1)$ to $4k$, $(\xb \cdot B_k \xb)$ denotes the scalar product between the vectors $\xb$ and $B_k \xb$, the $c_\a$ are functions of the quantities $(\xb \cdot B_k \xb)$ ($k=1,...n$); and the $L_\a$ are block-diagonal matrices with four-dimensional blocks, satisfying the quaternionic relations. These are now written in the form $$ L_\a \, L_\b \ = \ \sum_\ga \, \epsilon_{\a \b \ga} \, L_\ga - \de_{\a \b} \, \E_{4n} \ , $$
with $\E_{4n}$ the $4n$-dimensional identity matrix. \EOR

\medskip\noindent
{\bf Remark 8.} Writing, with an obvious notation, the $4n$-dimensional (and block-reducible) matrices $L_\a$ as $$ L_\a \ = \ L_\a^{(1)} \oplus ... \oplus L_\a^{(n)} \ , $$ it is immediate to see that the submatrices corresponding to each four-dimensional block also satisfy the quaternionic relations \eqref{eq:quaternionic}. By Remark 3 we conclude that acting on $\R^{4n}$ via a linear transformation in $\mathrm{SO(4)} \times ... \times \mathrm{SO(4)} \ss \mathrm{SO}(4n)$, in each block the matrices $L_\a^{(k)}$ can be reduced to standard ones, with either positive or negative orientation depending on the matrices $L_\a$, say with $m=0,...,n$ positively oriented and $n-m$ negatively oriented blocks. Note also that by a (block) permutation of variables (which does not alter the orientation in $\R^{4n}$), we can always reduce to the case where the first $m$ blocks have positive orientation, and the remaining $(n-m)$ have negative one. \EOR

\subsection{Dynamics of quaternionic oscillators}

We will now consider the dynamics of quaternionic oscillators. We will first consider the situation in the simplest possible case (i.e. dimension four, a single quaternionic oscillator) and then more general cases.

\subsubsection{Simple quaternionic oscillators}
\label{sec:simpleQO}

Let us consider the dynamics of simple quaternionic oscillators, \eqref{eq:simplequatosc}. We rewrite the equations as
\beq \label{eq:quatoscL} \xd^i \ = \ L^i_{\ j} \ x^j \ , \ \ \ L \ = \ \sum_\a c_\a (|\xb|^2) \, L_\a \ . \eeq
It should be noted that $\rho := |\xb|^2$ is always a constant of motion for such a dynamics. In fact, recalling that the $L_\a$, and hence $L$, are skew-symmetric, we have
\beq \label{eq:rho} \frac{d \rho}{d t} \ = \ 2 \ x_i \ \dot{x}^i \ = \ 2 \ x_i \ L^i_{\ j} \, x^j \ = \ 0 \ . \eeq
This means that $\rho$, and hence the $c_\a (\rho)$ and the matrix $L = L (\a) = \sum_\a c_\a (\rho) L_\a$, can be considered as constant on the dynamics.
In view of this remark, it is clear that the solution to the system \eqref{eq:quatoscL} is
\beq \label{eq:solsqo} \xb (t) \ = \ \exp[ L t] \ \xb (0) \ ; \eeq
As for $L^2$, we have
\begin{eqnarray*}
L^2 &=& \( \sum_\a c_\a \, L_\a \) \ \( \sum_\b c_\b L_\b \)
= \sum_{\a,\b}  c_\a \, c_\b \ \[ \epsilon_{\a \b \ga} \, L_\ga \ - \ \de_{\a \b} \E_{4} \] \\
&=& - \ \( \sum_\a c_\a^2 \) \ \E_4 \ := \ - \ \nu^2 \ \E_{4} \ . \end{eqnarray*}

Now we note that $\nu^2$ depends only on the $c_\a$, and hence it depends on the $x^i$ only through $\rho$. Again by \eqref{eq:rho}, it follows that $\nu^2$, and hence $\nu = \sqrt{\nu^2}$, are constants of motion.

We can now go back to \eqref{eq:solsqo}: due to $\nu^2$ being a constant and taking also into account $L^2 = - \nu^2 \E_{4}$, we have, recalling the series expansion for $e^{Lt}$, that
\beq \label{eq:qoscsol1} \exp[L t] \ = \ \sin (\nu t) \, L \ + \ \cos (\nu t) \, \E_{4} \ . \eeq The solution \eqref{eq:solsqo} is therefore, with $\xb_0 = \xb (0)$ the initial condition, written as
\beq \label{eq:solsqo2} x(t) \ = \ \[ \cos (\nu t) \ \E_{4} \ + \ \sin (\nu t) \ L \] \  \xb_0 \ . \eeq

The solutions live on the sphere $S^3$ of radius $|\xb_0|$ (as already apparent from $d \rho / dt = 0$), moving on great circles $S^1$ identified by $\xb_0$ and $\xb_1 = L \xb_0$. They realize the Hopf fibration of $S^3$ \cite{Hus}. We stress that -- albeit we have not written this explicitly to avoid a heavy notation -- in \eqref{eq:qoscsol1} and \eqref{eq:solsqo2} $\nu$ is a function of $\rho$; thus it is a constant of motion, but takes in general different values on different spheres.

Note that if the system is Hamiltonian we actually have two global constants of motion (e.g., if the only nonzero Hamiltonian is $\h_1$ and $L_\a = Y_\a$, these are $I_1 = (x^1)^2 + (x^2)^2$ and  $I_2 = (x^3)^2 + (x^4)^2$), while for genuinely \hh systems we have only $\rho = |\xb|^2$.\footnote{Needless to say, as we have a closed curve in a four-dimensional space, we always have three constants of motion; the point is that the other two will depend on the radius $\rho$ of the sphere, or more precisely on the values taken by $c_\a (\rho)$ on these spheres; see Remark 10 below for a more precise statement.}

The equivalent of action-angle coordinates are now {\it action-spin} coordinates $(I, s^\a)$, where now $I = |\xb|^2 \in \R$, and the $s^\a$ are coordinates in $\mathrm{SU(2)} \simeq S^3$ (as the sphere $S^3$ is parallelizable \cite{Hus}, these are global coordinates). This also shows that quaternionic oscillators describe an evolution on the $\mathrm{SU(2)}$ group, governed by an element of the $\mathrm{su(2)}$ Lie algebra which depends only on $|\xb|^2 = I$ and is hence constant on the level manifolds for $I$ (i.e. on spheres $S^3$ of given radius).

\medskip\noindent
{\bf Remark 9.} It should be stressed that while in Hamiltonian dynamics each constant of motion allows to reduce one degree of freedom, i.e. to lower the dimension of the system of first order ODEs by two, for genuinely \hh systems each constant of motion still allows to reduce one (quaternionic) degree of freedom, but now this means lowering the dimension of the system of first order ODEs by four. \EOR

\medskip\noindent
{\bf Remark 10.} As mentioned above, integrability implies we have closed curves as trajectories of solutions, and hence (being in $\R^4$) three constants of motion. By explicit computations, one finds out that the two additional constants of motion -- beside $\rho = |\xb|^2 = Q_1$ -- can be chosen as
\begin{eqnarray*}
Q_2 &=& c_1 \, \( x_1^2 + x_2^2 \) \ + \ c_2 \, \( x_2 x_4 - x_1 x_3 \) \ + \ c_3 \, \( x_1 x_4 + x_2 x_3 \) \ , \\
Q_3 &=& c_3 \, \( x_1^2 + x_3^2 \) \ + \ c_1 \, \( x_2 x_3 - x_1 x_4 \) \ + \ c_2 \, \( x_1 x_2 + x_3 x_4 \) \ . \end{eqnarray*}
Actually we could choose different ones as well; in fact, the quantities
\begin{eqnarray*}
B_{12} &=& c_1  (x_1^2+x_2^2 ) + c_3 (x_2  x_3 + x_1  x_4  ) + c_2  (x_2  x_4 - x_1  x_3 ) \\
B_{13} &=& c_3  (x_1^2+x_3^2 ) + c_2 (x_1  x_2 + x_3  x_4  ) + c_1  (x_2  x_3 - x_1  x_4 ) \\
B_{14} &=& c_2  (x_1^2+x_4^2 ) + c_1 (x_1  x_3 + x_2  x_4  ) - c_3  (x_1  x_2 - x_3  x_4 ) \\
B_{23} &=& c_2  (x_2^2+x_3^2 ) - c_1 (x_1  x_3 + x_2  x_4  ) + c_3  (x_1  x_2 - x_3  x_4 ) \\
B_{24} &=& c_3  (x_2^2+x_4^2 ) + c_1 (x_1  x_4 - x_2  x_3  ) - c_2  (x_1  x_2 + x_3  x_4 ) \\
B_{34} &=& c_1  (x_3^2+x_4^2 ) - c_3 (x_2  x_3 + x_1  x_4  ) + c_2  (x_1  x_3 - x_2  x_4 )
\end{eqnarray*}
are all invariant ones; but only three of them are functionally independent. \EOR

\def\ri{\mathtt{i}}
\def\rj{\mathtt{j}}
\def\rk{\mathtt{k}}
\def\rd{\mathtt{d}}

\medskip\noindent
{\bf Remark 11.} In this note we have chosen to use a real notation for quaternions; one could introduce
quaternionic imaginary units $\ri,\rj,\rk$ satisfying
$$ \ri \rj \, = \, \rk \ , \ \  \ri^2 \, =\ , \rj^2 \, = \, \rk^2 \, = \, -1 \ , $$
and cyclic permutations; and the quaternionic variable
$$ q \ = \ x_1 \ + \ \ri \, x_2 \ + \ \rj \, x_3 \ + \ \rk \, x_4 \ ; \ \
\bar{q}\ = \ x_1 \ - \ x_2 \, \ri \ - \ x_3 \, \rj \ - \ x_4 \, \rk \ . $$
In this notation, and introducing the (pure imaginary) quaternionic Hamiltonian
\beq \mathtt{H} \ = \ \ri \, \h_1 \ + \ \rk \, \h_2 \ + \ \rj \, \h_3 \ , \eeq
the equations of motion \eqref{eq:quatoscL} for the simple quaternionic oscillator read
\begin{eqnarray*}
\dot{q} &=& \dot{x}_1 + \ri \dot{x}_2 + \rj \dot{x}_3 + \rk \dot{x}_4 \ = \
- \left(
  \frac{\pa \mathtt{H}}{\pa x_1}
+ \frac{\pa \mathtt{H}}{\pa x_2} \ri
+ \frac{\pa \mathtt{H}}{\pa x_3} \rj
+ \frac{\pa \mathtt{H}}{\pa x_4} \rk
\right) \ .
\end{eqnarray*}

If now we introduce $\vec{e}=(\ri,\rj,\rk)$ and define, following \cite{Sud}, the operators
\begin{eqnarray*}
\partial_r f &:=& \frac12 \ \left( \partial_{x_1} \, f \ - \ (\nabla f) \, \cdot \, \vec{e} \right)
\ , \ \ \bar{\partial}_r f \ := \frac12 \ \left( \partial_{x_1} \, f \ +\ (\nabla f) \, \cdot \, \vec{e} \right)
\ ,\\
\partial_\ell f &:=& \frac12 \ \left( \partial_{x_1} \, f \ - \ \vec{e} \, \cdot \, (\nabla f) \right)
\ , \ \ \bar{\partial}_l f \ := \ \frac12\ \left( \partial_{x_1} \, f \ + \ \vec{e} \, \cdot \, (\nabla f) \right) \ ,
\end{eqnarray*}
then the equations can be written as
$\dot{q} = - 2 \bar{\partial}_r \mathtt{H}$.
Note that as $\overline{ab}=\bar{b}\bar{a}$,  the conjugate equation is
$\dot{\bar{q}} = - 2 \partial_\ell \bar{\mathtt{H}}$.

In this notation, $ \rho = |q|^2 = q \bar{q}$, and it immediate to check that the above equations guarantee
it is a constant of motion. In fact,
\begin{eqnarray*}
\frac{d \rho}{d t} &=& \dot{q} \, \bar{q} \ + \ q \, \dot{\bar{q}} \ = \ - 2 \, \( \bar{\partial}_r  \mathtt{H} \) \, \bar{q} \ - \ 2 \, q \, \( \partial_\ell  \bar{\mathtt{H}} \) \ = \
- 2 \, \[ \( \bar{\partial}_r  \mathtt{H} \) \, \bar{q} \ -\ q \, \( \partial_\ell   \mathtt{H} \) \] \ ;
\end{eqnarray*}
but, as
$ \mathtt{H}_{\alpha} = \h_\a =  (1/2) c_{\alpha} q \bar{q}$, and $\bar{c}=-c$,
we easily get
\begin{eqnarray*}
\bar{\partial}_r \mathtt{H} \ = \ ( \bar{\partial}_r \mathcal{H}_1) \, \ri \ + \ (\bar{\partial}_r \mathcal{H}_2) \, \rk \ + \ (\bar{\partial}_r \mathcal{H}_3) \, \rj \ =\ \frac12 \ q \ \mathbf{c} \ , \\
\partial_\ell \mathtt{H} \ = \ (\partial_\ell \mathcal{H}_1) \, \ri \ + \ (\partial_\ell \mathcal{H}_2) \, \rk \ + \ (\partial_\ell \mathcal{H}_3) \, \rj \ = \ \frac12 \ \mathbf{c}\ \bar{q} \ ,
\end{eqnarray*}
where we have written for short
$ \mathbf{c} := c_1 \ri + c_2 \rk + c_3 \rj $.
The conclusion $d \rho / d t = 0$
follows immediately. \EOR

\subsubsection{Generic quaternionic oscillators}
\label{sec:genericQO}

For generic quaternionic oscillators, the discussion goes pretty much the same. In fact, each $\rho_k = |\xi_{(k)} |^2$ is a constant of motion (as follows again from $L^T = - L$), so that the $c_\a$ in \eqref{eq:quatosc} are constant on the dynamics. The matrices $L_\a$ are block-reducible, $L_\a = L_\a^{(1)} \oplus ... \oplus L_\a^{(k)}$. Thus
$$ \exp[ t L_\a ] \ = \ \exp [t L_\a^{(n)} ] \oplus ... \oplus \exp [t L_\a^{(n)} ] \ , $$ and each $\exp [t L_\a^{(k)} ]$ is computed in the same way as shown above for simple quaternionic oscillators.

We thus reach the same conclusion for the solution issued from an initial datum $x(0) = x_0$ (now with $x_0 \in \R^{4 n}$), i.e. it is given by
\beq \label{eq:solgqo2} x(t) \ = \ \[ \cos (\nu t) \ \E_{4n} \ + \ \sin (\nu t) \ L \] \  x_0 \ . \eeq
Now we have action-spin coordinates $({\bf I},{\bf s}^\a)$ with ${\bf I} = (I_{(1)},...,I_{(n)}) $, and similarly ${\bf s}^\a =   ( s^\a_{(1)} , ... , s^\a_{(n)} )$. Note that now we may have a different set of matrices (i.e. a different $\mathrm{su(2)}$ representation) in each block.

Here the actions $I_{(k)}$ are constants of motion; with the notation used in section \ref{sec:introQO}, we have $I_{(k)} = | \xi_{(k)} |^2$.

\subsection{Dirac systems and Dirac oscillators}
\label{sec:dirac}

We have so far considered systems related to a given \hk structure. However, as mentioned in Remark 5 (and as discussed in detail in \cite{GRcanonical}), one could have system related to a given {\it pair of dual \hk structures}; this is the case e.g. for the \hh description of the Dirac equation, and hence one speaks of {\it Dirac systems}.

As mentioned in Remark 5, Dirac systems, i.e. vector fields
\beq X \ = \ X_{(+)} \ + \ X_{(-)} \eeq which decompose as the sum of vector fields $X_{(\pm)}$ which are \hh with respect to a pair of dual (positively and negatively oriented) \hk structures can be dealt with through Walcher factorization principle \cite{Wal}. Now we want to consider this kind of situation, when $X_{(\pm)}$ correspond to quaternionic oscillators; we will refer to this case as {\it Dirac oscillators}. We stress they represent a generalization of quaternionic oscillators, and have not been studied in previous works on quaternionic oscillators and \hh integrable systems.

We will work directly with the standard \hk structures; that is, we will have
\beq\label{eq:Xdirac} X_{(\pm)} \ = \ f^i_{(\pm)} (x) \ \pa_i \ , \eeq
with coefficients $f^i_{(\pm)}$ given by
\beq
f^i_{(+)} (x) \ = \ \sum_{\a = 1}^3 c_\a (\rho) \ (Y_\a)^i_{\ j} \, x^j \ , \ \
f^i_{(-)} (x) \ = \ \sum_{\a = 1}^3 \^c_\a (\rho) \ (\^Y_\a)^i_{\ j} \, x^j \ ; \eeq
here $\rho = x_1^2 + ... + x_4^2 = |\xb |^2$.
It will be convenient to write
$$ \nu_{(+)} (\rho) \ := \ \sqrt{c_1^2 (\rho) + c_2^2 (\rho) + c_3^2 (\rho) } \ , \ \ \nu_{(-)} (\rho) \ := \ \sqrt{\^c_1^2 (\rho) + \^c_2^2 (\rho) + \^c_3^2 (\rho) } \ ; $$ and moreover $$ K_{(+)} \ = \ \frac{1}{\nu_{(+)} (\rho) } \ \sum_{\a = 1}^3 c_\a (\rho) \ (Y_\a)^i_{\ j} \ ; \ \
K_{(-)} \ = \ \frac{1}{\nu_{(-)} (\rho) } \ \sum_{\a = 1}^3 \^c_\a (\rho) \ (\^Y_\a)^i_{\ j} \ . $$
Note that $[Y_\a, \^Y_\b] = 0$ for all $\a,\b$ immediately implies $[K_{(+)} , K_{(-)} ] = 0$.
We will also consider $$ K \ = \ K_{(+)} \ + \ K_{(-)} \ . $$

The dynamic under $X$ [and that under $X_{(\pm)}$] will then be described by, respectively,
$$ \xd \ = \ K \, x \ , \ \ \ \[ \mathrm{and} \ \ \xd \ = \ K_{(\pm)} \, x \ \] \ . $$
We know that, as shown above, the dynamics under $X_{(\pm)}$ can be integrated. We will show that the dynamics under $X$ is also explicitly integrable

Note that as $Y_\a^T = - Y_\a$ and $\^Y_\a^T = - \^Y_\a$, the matrices $K_{(\pm)}$ and $K$ are all antisymmetric. It follows at once that on the dynamics $\rho$ is constant. This also means that for any initial datum $\xb (0)$, we can consider the $c_\a$ and $\^c_\a$ coefficients (which depend on $x$ only through $\rho$) as constant. The same holds for $\nu_{(\pm)} (\rho)$. We will thus from now on just omit to indicate their dependence on $\rho$ when dealing with a single solution.

Let us now look at the flow under $X$ as in \eqref{eq:Xdirac}; as mentioned in Remark 5, Walcher's factorization principle \cite{Wal} states that denoting by $\Phi(t;x_0;Y)$ the time $t$ flow issuing from $x_0$ at time $t=0$ under the vector field $Y$, we have
\beq\label{eq:flowX} \Phi(t;x_0;X) \ = \ \Phi\[t; \Phi(t;x_0;X_{(+)} ); X_{(-)} \] \ = \ \Phi\[t; \Phi(t;x_0;X_{(-)} ); X_{(+)} \] \ . \eeq
In the case we are presently considering, the flows under $X_{(\pm)}$ can be explicitly computed, see Sect.\ref{sec:simpleQO}; more precisely, with the present notation, we have
\beq
\Phi(t;\xb_0;X_{(\pm)} ) \ = \ \[ \cos (\nu_{(\pm)} t) \, \E_4 \ + \ \sin( \nu_{(\pm)} t) \, K_{(\pm)} \] \ \xb_0 \ := \ A_{(\pm)} \ \xb_0 \ . \eeq
Needless to say, $[K_{(+)} , K_{(-)} ] = 0$ entails $[A_{(+)} , A_{(-)} ] = 0$ as well.

With the present notation, and writing for short
$$ \chi_\pm \ = \ \cos (\nu_{(\pm)} t) \ , \ \ \s_\pm \ = \ \sin (\nu_{(\pm)} t) \ , $$
the matrices $A_{(\pm)}$ are given in explicit terms by
{\small
\begin{eqnarray}
A_{(+)} &=& \pmatrix{
  \chi_+ & c_1 \s_+ & c_3 \s_+ & c_2 \s_+ \cr
- c_1 \s_+ & \chi_+ & c_2 \s_+ & - c_3 \s_+ \cr
- c_3 \s_+ & - c_2 \s_+ & \chi_+ & c_1 \s_+ \cr
- c_2 \s_+ & c_3 \s_+ & - c_1 \s_+ & \chi_+ \cr} \ ; \label{eq:A+} \\
A_{(-)} &=& \pmatrix{
  \chi_- & - \^c_3 \s_-  & \^c_1 \s_- & - \^c_2 \s_- \cr
 \^c_3 \s_- & \chi_- & \^c_2 \s_- & \^c_1 \s_- \cr
- \^c_1 \s_- & - \^c_2 \s_- & \chi_- & \^c_3 \s_- \cr
 \^c_2 \s_- & - \^c_1 \s_- & - \^c_3 \s_- & \chi_- \cr}
 \ . \label{eq:A-} \end{eqnarray}
}

According to \eqref{eq:flowX}, the flow under $X$ will then be described by
\beq \Phi(t;\xb_0;X) \ = \ A_{(+)} (t) \, A_{(-)} (t) \ \xb \ = \ A_{(-)} (t) \, A_{(+)} (t) \ \xb \ := \ A (t) \, \xb ; \eeq
this matrix $A(t)$ can also be written as
\begin{eqnarray}
A (t) &=& [\cos (\nu_{(+)} t) \, \cos (\nu_{(-)} t)] \ I \ + \ [\cos (\nu_{(+)} t) \, \sin (\nu_{(-)} t)] \ K_{(+)} \\ & & + \ [\sin (\nu_{(+)} t) \, \cos (\nu_{(-)} t)] \ K_{(-)} \ + \ [\sin (\nu_{(+)} t) \, \sin (\nu_{(-)} t)] \ K_{(+)} \, K_{(-)} \ . \nonumber \end{eqnarray}

A more explicit expression is immediately obtained by multiplying the two matrices $A_{(\pm)}$ as given in \eqref{eq:A+} and \eqref{eq:A-}; this is long and not specially interesting and hence will not be reported here.

It is a matter of straightforward -- albeit rather boring -- algebra to check that indeed
$\xb (t) = A(t) \xb_0 $ is a solution to $d \xb / dt = X ( \xb) = K \xb$, for any initial condition
$\xb (0) = \xb_0$; this is also seen by simply checking that $ (d A/d t) = K A$.

%

\subsection{Asymptotically integrable Dirac and quaternionic \\ systems}

It is quite remarkable that systems associated to quaternionic or Dirac oscillators via adding a
gradient vector field present the phenomenon of {\it spontaneous linearization} \cite{CGspo}, which in
this context means {\it asymptotic integrability}.

In fact, let us consider a system of the type
\beq \xd^i \ = \ f_0 (|\xb|^2) x^i \ + \ \sum_\a c_\a (|\xb|^2) \, (Y_\a)^i_{\ j} x^j \ + \ \ + \ \sum_\a \^c_\a (|\xb|^2) \, (\^Y_\a)^i_{\ j} x^j \ ; \eeq
then the dynamics of $\rho = |\xb|^2$ is controlled by $f_0$ alone (due to $Y_\a^T = - Y_\a$, $\^Y_\a^T = - \^Y_\a$),
\beq \frac{d \rho}{dt} \ = \ 2 \ f_0 (|\xb|^2 ) \ |\xb|^2 \ . \eeq Thus it will evolve towards the stable zeros of $f_0$ (those with $f' (\rho_0) < 0$). On spheres with such radius, which are reached asymptotically by the dynamics, the system will behave as a Dirac oscillator -- or a quaternionic one if only the $c_\a$ or only the $\^c_\a$ are nonzero -- and hence is integrable.

\section{Symmetry of integrable hyperhamiltonian \\ systems}
\label{sec:symmint}

We do now want to discuss the symmetry properties of integrable \hh systems; here we will consider the special \hh systems which are Hamiltonian as degenerate systems, and focus instead on the generic case of non-Hamiltonian integrable \hh systems (see \cite{Gdeg} for a discussion).

Let us define more precisely the notion of symmetry of a dynamical system. A dynamical system $\xd = f(x)$ on a manifold $M$ is characterized by the vector field $X = f^i (x) \pa_i$. If we consider a map $\Phi : M \to M$, this induces a (push-forward) map $\Phi_*$ on $\T M$, hence on the vector fields on $M$; if this satisfies $\Phi_* (X) = X$, we say that $\Phi$ is a symmetry of (the dynamical system characterized by) $X$. It is customary to express this notion by saying that $\Phi$ preserves the form of the equation $\xd = f(x)$. In many cases one is interested in one-parameter (or multi-parameter) families of maps $\Phi_\a : M \to M$, generated by a vector field $Z : M \to \T M$. In this case we say, with a slight abuse of language, that $Z$ is a symmetry of $X$ if the one-parameter group $\Phi_\a$ generated by $Z$ is a group of symmetries of $X$ (in some cases, we are satisfied with a local group, i.e. it suffices that $\Phi_\a$ is defined for $\a$ in a neighborhood of zero, where $\Phi_0$ is the identity map).

The symmetry properties of a dynamical system are conserved under diffeomorphisms. Thus, as quaternionic integrable systems are characterized by the property of being diffeomorphic to a system of oscillators \eqref{eq:quatosc}, \eqref{eq:quatoscbis}, we can investigate the symmetry properties of quaternionic integrable systems by working directly on \eqref{eq:quatosc}, or equivalently on \eqref{eq:quatoscbis}.

Let us consider the simple quaternionic oscillator in $\R^4$, with equations given by \eqref{eq:quatosc} and study the symmetries of the system, that is, the transformations leaving invariant the form of these equations; the hypothesis that the system is not Hamiltonian does in this framework imply that the $c_\a$ do actually depend on the $\xi_\a$, i.e. that the matrix $L (\rho)$ describing quaternionic oscillators dynamics is not constant. We will restrict in the sequel to linear transformations of the coordinates\footnote{This restriction is actually not needed: if we denote by $\mathcal{G}_0$ the Lie algebra of linear symmetries, and by $\mathcal{G}$ that of general Lie-point symmetries, it turns out that $\mathcal{G} = \mathcal{I} \otimes \mathcal{G}_0$, where $\mathcal{I}$ is the ring of smooth constants of motions, i.e. in this case the ring of smooth functions of $\rho$.} (sum over repeated latin indices is understood):
\beq
x'^i=\Lambda^i_{\hphantom{i}j}x^j
\eeq
where $\Lambda$ is a constant regular matrix. Since, after the discussion in Section 3.1, the trajectories are confined to the sphere with constant $\sum_i|x^i|^2=\rho^2$, the matrix $\Lambda$ should be orthogonal\footnote{This follows from the requirement that the $c_\a$ and hence the matrix $L$ do effectively depend on $\rho$ (see above); should these be actually constant, a dilation would (commute with time evolution and hence) be a symmetry.}. In fact, we will consider $\Lambda \in \mathrm{SO(4)}$ if we want to keep the orientation unchanged. Substituting in the equations of motion \eqref{eq:quatoscL} we easily get:
\beq
\dot{x}'^i=\Lambda^i_{\hphantom{i}j}\dot{x}^j=\sum_\alpha c_{\alpha}\Lambda^i_{\hphantom{i}j}(L_{\alpha})^{j}_{\hphantom{j}k}(\Lambda^{-1})^k_{\hphantom{k}l}x'^{l}
\eeq
and the system is invariant if
\beq\label{first}
\Lambda \ L_{\alpha} \ \Lambda^{-1} \ = \ L_{\alpha} \ , \ \ \alpha=1,2,3 \ .
\eeq

Albeit the above discussion has been made in $\R^4$, the extension to any $\R^{4n}$ space is straightforward, once the matrices $L_{\alpha}$ have been block-diagonalized (in $4\times 4$ blocks, see Remark 8). The problem we have to solve is to find the group of linear transformations satisfying the equations:
\beq
\Lambda \, \Lambda^t \ = \ \E_4 \ , \ \  \det\Lambda \ = \ 1 \ , \ \ \Lambda \, L_{\alpha} \, \Lambda^{-1} \ = \ L_{\alpha} \ , \ \  \alpha=1,2,3.
\eeq	
This group is obviously a subgroup of the orthogonal group and the computation is easy to do in $\R^4$. In fact, since the matrices $L_{\alpha}$ satisfy the quaternionic relations, they generate a linear representation (of dimension 4) of the Lie algebra $\mathrm{su}(2)$ and equation \eqref{first} becomes
\beq
\Lambda \, L_{\alpha} \ = \ L_{\alpha} \, \Lambda \ , \ \  \alpha=1,2,3.
\eeq
Since the representation is real irreducible, but complex reducible, the (real) Schur lemma (see e.g., chapter 8 of \cite{Kir}) implies that $\Lambda$ is an element of the subgroup of $\mathrm{SO}(4)$ generated by the matrices of the complementary structure to that generated by the matrices $L_{\alpha}$, that is, $\mathrm{SO}(3)$ (since $\mathrm{SO}(4)\sim \mathrm{SO}(3)\times \mathrm{SO}(3)$).

Equation \eqref{first} is a sufficient condition to assure the invariance of the system, but we can also consider the more general condition (which obviously contains \eqref{first} as a particular case)
\beq
\sum_\alpha c_{\alpha} \ \Lambda \, L_{\alpha} \, \Lambda^{-1} \ = \ \sum_\alpha \ c_{\alpha} \, L_{\alpha}
\eeq
Since the matrices $L_{\alpha}$ satisfy the quaternionic relations, we will assume that the matrices transformed under $\Lambda$ are a linear combination of the original ones, and satisfy the quaternionic relations:
\beq\label{inva}
\Lambda \ L_{\alpha} \ \Lambda^{-1} \ = \ \sum_{\beta} \, R_{\alpha\beta} \, L_{\beta} \ .
\eeq
As it is easy to prove, the matrix $R$ is necessarily a rotation in $\R^3$ and
the invariance condition is
\beq
\sum_{\beta} c_{\alpha} R_{\alpha\beta} =  c_{\alpha}
\eeq
which implies that the vector $c_{\alpha}$ is an eigenvector of the rotation $R$ (or its inverse), or equivalently, the matrix $R$ is a rotation with a fixed axis, and then, the group generated by $R$ is isomorphic to $\mathrm{SO}(2)$.

In order to determine the possible matrices $\Lambda$ satisfying this equation, we can approach the problem from an infinitesimal point of view. At first order in $\varepsilon$, we have
\beq
\Lambda=I_{4n}+\varepsilon X,\quad X+X^T=0,\quad R=I_3+\varepsilon {\cal J},\quad {\cal J}+{\cal J}^T=0 \ .
\eeq
where ${\cal J}$ is the generator of the uniparametric group of rotations with axis $c_{\alpha}$, that is
\beq\label{skew}
\sum_{\beta}{\cal J}_{\alpha\beta}c_\beta=0
\eeq
Then, the invariance equation (\ref{inva}) is transformed into:
\beq \label{trans} [X,L_{\alpha}]= \sum_{\beta=1}^3
{\cal J}_{\alpha\beta}L_{\beta},\quad \alpha=1,2,3.
\eeq
This is the infinitesimal invariance equation, which will be the main tool to determine $X$, i.e., $\Lambda$.

The computation of the invariance groups is an easy task in the cases we are considering, $(\R^{4n},I_{4n})$. In fact, if $n=1$, we can write the skew-symmetric matrix $X\in \mathrm{so}(4)$ as a linear combination of the quaternionic matrices $L_{\alpha}$ and $\widehat{L}_{\alpha}$ corresponding to the two $\mathrm{su}(2)$ algebras in the decomposition $\mathrm{so}(4)=\mathrm{su}(2)\oplus \mathrm{su}(2)$:
 \beq X \ = \ \frac12 \, \sum_{\beta=1}^3 \, a_{\beta} \, L_{\beta} \ + \ \frac12 \, \sum_{\beta=1}^3 \, \widehat{a}_{\beta} \, \widehat{L}_{\beta}, \quad [L_{\alpha},\widehat{L}_{\beta}]=0
 \eeq
The invariance equation for the positively oriented standard structure $L_{\beta}$ yields the solution
\beq \label{defM}
{\cal J}_{\alpha\beta} \ = \ \sum_{\gamma=1}^3 \, \epsilon_{\alpha\beta\gamma} \, a_{\gamma} \ , \ \ \ \ \alpha,\beta=1,2,3. \eeq
Since ${\cal J}$ satisfies  equation \eqref{skew}, the constant vector $a_{\alpha}$ should satisfy
\beq
\sum_{\gamma} \, \epsilon_{\alpha\beta\gamma}c_{\beta}a_{\gamma}=0
\eeq
that is, the vector $a_{\alpha}$ and $c_{\beta}$ are proportional and the invariance group has as infinitesimal generator
\beq\label{gen}
{\cal J}_{\alpha\beta}=\sum_{\gamma}\epsilon_{\alpha\beta\gamma}c_{\gamma}
\eeq
To summarize, the whole invariance algebra for the quaternionic oscillator equations
\eqref{eq:quatoscL} in dimension 4 is \beq\label{eq:L1} \mathcal{L}_1 \ = \ \mathrm{so}(2) \oplus \mathrm{su}(2) \eeq and the generator of the algebra $\mathrm{so}(2)$ is determined by the constants $c_{\alpha}$, see \eqref{gen}.

Let us now consider $\R^8$. The matrices $L_{\alpha}$ are a quaternionic structure, and we will assume that they are written as $4\times 4$ diagonal block matrices\footnote{The resulting invariance group will be conjugated to that of an arbitrary representation of the quaternionic structure.}, each block having a definite orientation. It is easy to show that all of these structures are conjugated under $\mathrm{O}(8)$ and then we can reduce the study to that of a positive oriented quaternionic structure:
\beq
L_{\alpha}=\left(\begin{array}{cc} \Y_{\alpha} & 0 \\ 0& \Y_{\alpha}\end{array}\right)
\eeq
We can easily show that the symmetry algebra of the quaternionic harmonic oscillator in $\R^8$ is
\beq\label{eq:L2} \mathcal{L}_2 \ = \ \mathrm{so}(2) \oplus \mathrm{sp}(2) \ ; \eeq
here and in the following $\mathrm{sp}(n)$ is the Lie algebra of the group $\mathrm{Sp}(n)$ of unitary four-dimensional symplectic matrices.\footnote{As some ambiguity is present in the literature concerning the notation for symplectic groups, it may be worth stating explicitly that for us the group $\mathrm{Sp}(n)$ will be the set of $2n \times 2n$ (complex) unitary symplectic matrices (thus with real representation of dimension $4n$), with Lie algebra $\mathrm{sp}(n) \ss \mathit{Mat}(2n;{\bf C}) \simeq \mathit{Mat}(4n;\R)$.}

We will just sketch the proof. The main idea is to construct, as in $\R^4$, a basis of $\mathrm{o}(8)$ starting from the quaternionic matrices $\Y_{\alpha}$ and $\widehat\Y_{\alpha}$. In fact, we need some symmetric $4\times 4$ matrices to complete the basis, which can be written as
\begin{eqnarray}
\left(\begin{array}{cc} \Y_{\alpha} & 0 \\ 0& 0
\end{array}\right),\; \left(\begin{array}{cc} 0 & 0 \\ 0&  \Y_{\alpha}
\end{array}\right),\; \left(\begin{array}{cc}  \widehat{\Y}_{\alpha} & 0 \\ 0& 0
\end{array}\right),\; \left(\begin{array}{cc} 0 & 0 \\ 0& \widehat{\Y}_{\alpha}
\end{array}\right)\nonumber \\
\left(\begin{array}{cc} 0 & \Y_{\alpha} \\ \Y_{\alpha}& 0
\end{array}\right),\; \left(\begin{array}{cc} 0 & \widehat{\Y}_{\alpha} \\ \widehat{\Y}_{\alpha}& 0
\end{array}\right),\; \left(\begin{array}{cc} 0 & S_i \\ -S_i& 0
\end{array}\right).
\end{eqnarray}
with $\alpha=1,2,3$, and $S_i$, $i=1,\ldots,10$, the set of $4\times 4$ elementary symmetric matrices (that is, $E_{jj}$ and $E_{jk}+E_{kj}$, where $E_{jk}$ is the elementary matrix with 1 in the position $jk$ and $0$ elsewhere).

Applying the equations (\ref{trans}), we get the matrices $X$
\beq\label{basis}
\left(\begin{array}{cc} \widehat\Y_{\alpha} & 0 \\ 0& 0\end{array}\right),\;
\left(\begin{array}{cc} 0 & 0 \\ 0& \widehat\Y_{\alpha}\end{array}\right),\;
\left(\begin{array}{cc} 0 & \widehat\Y_{\alpha} \\ \widehat\Y_{\alpha}& 0\end{array}\right)
\left(\begin{array}{cc} 0 & \E_4 \\ - \E_4& 0\end{array}\right).
\eeq
The matrix ${\cal J}$ is
\beq\label{rot} \left(\begin{array}{cc} \widetilde{\cal J} & 0
\\ 0& \widetilde{\cal J}\end{array}\right), \eeq
where $\widetilde{\cal J}$ is the matrix we found in the 4-dimensional case. Then, it generates the algebra $\mathrm{so}(2)$. The other matrices generate a Lie algebra of dimension 10 which commutes with the above algebra $\mathrm{so}(2)$, and leaves invariant each of the matrices $L_{\alpha}$, $\alpha=1,2,3$.

It can be shown that these matrices are a representation of the symplectic algebra $\mathrm{sp}(2)$. Note that the case $\R^4$ has exactly the same structure, since $\mathrm{sp}(1) \approx \mathrm{su}(2)$.

The general case, i.e. quaternionic oscillators in $\R^{4n}$, is a generalization of the $8$-dimensional case and we get the general symmetry algebra:
\beq\label{eq:Ln}
\mathcal{L}_n \ = \ \mathrm{so}(2)\ \oplus \ \mathrm{sp}(n) \ .
\eeq

\medskip\noindent
{\bf Remark 12.} This fact is closely related to the computation of the holonomy group of hyperkahler and quaternionic manifolds, see for instance \cite{Sa89}. \EOR

\medskip\noindent
{\bf Remark 13.} Note that $\mathrm{su}(2) \oplus ... \oplus \mathrm{su}(2) \ss \mathrm{sp}(n)$; thus, as rather obvious, the symmetry algebra $\mathcal{L}_n$ include the product of $n$ independent $\mathrm{su}(2)$ algebras, each acting on one (quaternionic) degree of freedom. \EOR

\section{Integrable hyperhamiltonian systems and \\ symmetry}
\label{sec:intsymm}

In the previous section we have discussed and characterized the symmetry of a quaternionic integrable system. In this section we will reverse our point of view, i.e. discuss systems which enjoy the same symmetry properties of quaternionic integrable system.

In the Hamiltonian case (with compact energy manifolds), it is well known that a torus symmetry is enough to conclude that the system is integrable (see e.g. \cite{Arn1}); we want to investigate if something similar holds in the quaternionic case.

\medskip\noindent
{\bf Remark 14.} More precisely, in the Hamiltonian case (with compact energy manifolds), the symmetry is sufficient to fully characterize integrable systems. This is due to a topological lemma (see e.g. Lemma 2 of sect.49, p. 274, in \cite{Arn1}), which guarantees that if a compact connected $n$-dimensional manifold admits an abelian algebra of $n$ tangent vector fields spanning a regular $n$-dimensional distribution, then it is a torus ${\bf T}^n$.
We anticipate we are not able to provide a similar statement for the quaternionic case, i.e. to guarantee that a system with symmetry described by the algebra $\mathcal{L}_n$ is necessarily integrable, and this for the lack of a similar result for manifolds admitting an algebra of tangent vector fields corresponding to the $\mathcal{L}_n$ algebra. \EOR
\bigskip

First of all, we note that quaternionic oscillators in $n$ degree of freedom, i.e. in $\R^{4n}$, admit as invariant manifolds the product $$ V^n \ := \ S^3 \times ... \times S^3 \ \ \ (n \ \mathrm{factors}) \ , $$ where each $S^3$ factor belongs to an invariant $\R^4$ subspace; admitting such an invariant manifold is of course a necessary condition for a system to be a quaternionic oscillator.

Let us now consider the $\G = \mathrm{su}(2)$ algebra spanned by the $\^\Y_\a$ matrices, and look for vector fields $X = f^i \pa_i$ in $\R^4$ which are symmetric under this and do moreover leave the spheres $S^3$ invariant (i.e. admit $\rho$ as constant of motion); it is easy (e.g. by explicit computation) to check that these reduce to linear combinations -- with coefficients depending on $\rho$ -- of those associated to the $\Y_\a$. In other words, it results
\beq X \ = \ \sum_\a c_\a (\rho) \ (\Y_\a)^i_{\ j} x^j \ \pa_i \ . \eeq
Conversely, if we look at the $\G = \mathrm{su}(2)$ algebra spanned by the $\Y_\a$ matrices, and look for vector fields $\^X = f^i \pa_i$ in $\R^4$ which are symmetric under this and do moreover leave the spheres $S^3$ invariant, it results
\beq \^X \ = \ \sum_\a \^c_\a (\rho) \ (\^\Y_\a)^i_{\ j} x^j \ \pa_i \ . \eeq

Albeit this result is easily checked by explicit computation, its true nature follows from the Schur lemma in its real version: the only matrices which commute with the whole irreducible representation of $\G$ spanned by the $\^\Y_\a$ (respectively, by the $\Y_\a$) matrices, are the identity and those of the conjugated representation, i.e. the $\Y_\a$ (respectively, the $\^\Y_\a$).

This is then rephrased in terms of vector fields; the requirement to leave spheres invariant does of course exclude the dilation vector fields associated to the identity matrix. More precisely, the vector fields must be expressed in this way at each point; one could then think of coefficients $c_\a$ (respectively $\^c_\a$) being functions of $(x_1,...,x_4)$, but plugging such a vector field into the symmetry condition yields that they can actually only depend on the $x_i$ through $\rho = |x|^2$.  To make the argument completely clear, just consider the case where $\G$ is spanned by the $\^\Y_\b$ and thus the dynamical vector field is, as follows from the Schur Lemma argument, $X = \sum_\a c_\a (x) Y_\a$ (here we denote by $Y_\a$ the vector field $Y_\a = (\Y_\a)^i_{\ j} x^j \pa_i$ associated to the matrix $\Y_\a$, and similarly for $\^Y_\a$); then we immediately have that $[\^Y_\b , X] = \sum_\a \^Y_\b (c_\a) \, Y_\a$. As the $Y_\a$ are independent, this can vanish only if $\^Y_\b (c_\a) = 0$ for all $\a$ and $\b$, i.e. if the $c_\a$ only depends on joint invariants for the $\^Y_\b$; but the only function which is invariant under the three vector fields $\^Y_\b$ is $\rho = |x|^2$. The same holds, obviously, if we interchange the $\^\Y_\a$ and the $\Y_\a$.

Note that on each sphere $S^3$ the system will automatically have an $\mathrm{SO(2)}$ symmetry, corresponding to rotations around an axis identified by the $c_\a (\rho)$.

The general case (arbitrary $n$) is discussed along the same lines.  We recall Remark 13, and start considering the subalgebra $\G = \mathrm{su}(2) \oplus ... \oplus \mathrm{su}(2) \ss \L_n$ acting in $\R^{4n} = \R^4 \oplus ... \oplus \R^4$ leaving each $\R^4$ subspace invariant; it also preserves the $V_n = S^3 \times ... \times S^3$ manifolds, where each $S^3$ factor belongs to one of the $\R^4$ factors in the above mentioned splitting. The action of $\G$ in each $\R^4_{(m)}$ ($m=1,...,n$) invariant subspace is described by a linear combination of either the $\Y_\a$ or the $\^\Y_\a$ matrices; without any loss of generality, we can take as generators exactly either $L_\a^{(m)} = \Y_\a$ or $L_\a^{(m)} = \^\Y_\a$ (for each $m = 1,...,n$ we have either $\Y$ or $\^\Y$, but the same option is kept for all $\a=1,2,3$).

We thus look for vector fields in $\R^{4n}$ which are symmetric under $\G$. Proceeding as in the $n=1$ case, we first deal with matrices; writing this in four-dimensional block form, one easily concludes that elements on the block diagonal must commute with the corresponding $L_\a^{(m)}$ matrices, while those in off-diagonal blocks vanish (again this follows from the real version of Schur Lemma \cite{Kir}).

Thus in each block we are left with matrices in the conjugated representation of the $\mathrm{su}(2)$ Lie algebra. In other words, at each point we have a vector field associated to the conjugated representation of $\mathrm{SU}(2) \times ... \times \mathrm{SU}(2)$; when we pass to vector fields on $\R^{4n}$ we will have linear combinations of these with coefficients which  could in principles depend on the coordinates, but when we require the commutation with the symmetry vector fields and invariance of the $V_n = S^3 \times ... \times S^3$ manifolds, we are only left with the possibility of coefficients depending on $|\xi_1|^2 , ... , |\xi_n|^2$, for the same argument seen above in the $\R^4$ case.

We summarize our discussion as follows. We consider $\R^{4n}$ and the group $G = \mathrm{SU}(2) \times ... \times \mathrm{SU}(2)$ with each factor acting effectively on a subspace $\R^4$ of $\R^{4n}$ and trivially on the other ones, and with the action of $G$ in the $m$-th subspace $\R^4_{(m)}$ described by the $\mathrm{SU(2)}$ action generated by the $\Y_\a$ matrices (respectively by the $\^\Y_\a$ ones). The vector fields in $\R^{4n}$ which are symmetric under this group action are precisely those corresponding to quaternionic oscillators with suitable signature. These possess in turn additional symmetry as described by $\mathcal{L}_n$ identified in the previous Section. In other words, in this framework the symmetry properties characterize quaternionic integrable systems.

\medskip\noindent
{\bf Remark 15.} It should be stressed that we do {\it not } have an analogue of the result holding for Hamiltonian integrable systems. In fact, here we had to explicitly require invariance of the $V_n$ manifolds; on the other hand, in the Hamiltonian case the symmetry properties were sufficient to characterize the topology of the invariant manifolds (i.e. tori ${\bf T}^n$) and integrability followed from this. The point is that, to the best of our knowledge, one cannot state any correspondence between invariance under $\mathrm{SU(2)} \times ... \times \mathrm{SU(2)}$ and the topology of the manifolds $V_n = S^3 \times ... \times S^3$, contrary to what happens for $\mathrm{U(1)} \times ... \times \mathrm{U(1)}$ and ${\bf T}^n$. This entails, in particular, that our discussion -- conducted in the Euclidean framework -- cannot be extended to the general case as we are not able to control the global geometry in the general setting; in other words, we have a result which only holds locally on each chart. \EOR

\section{Discussion and conclusions}
\label{sec:concl}

We have discussed the notion of quaternionic integrable systems; these were defined as systems which can be mapped to a system of quaternionic oscillators, albeit later on we have seen that, thanks to Walcher's factorization principle, the scope of this definition can be enlarged to encompass Dirac oscillators as well. Physical applications of \hh dynamics, and in particular of quaternionic and Dirac oscillators, have been considered in previous papers \cite{GR,GRtnut}.

We have then characterized the symmetry properties of quaternionic oscillators; the required computations do actually to some extent reproduce those needed to study invariance properties of the quaternionic structure behind quaternionic oscillators, and have hence been only partially detailed here, referring to other works \cite{GRcanonical,GReuclidean} for details. As symmetries are invariant under diffeomorphisms, they are also properties of any quaternionic integrable system, and can be used to detect such systems. In the Euclidean case, suitable symmetry properties do characterize quaternionic integrable systems, but this does not extend to general (non-Euclidean) cases. Thus our general results concerning the relation between quaternionic integrability and symmetry properties provide a necessary (symmetry) condition for a system to be integrable, but not a sufficient one; this is due to the lack of an analogue of the ``topological lemma'' holding in the Hamiltonian case and guaranteeing any compact manifold with a torus action is actually a torus.

Finally, we would like to mention another aspect in which the quaternionic setting lacks an analogue of the familiar Hamiltonian one: in the case of general smooth symplectic manifolds, it is well known that there can be obstructions to global action-angle coordinates \cite{Dui}. As far as we know, there has been no study of how the Duistermaat results extend to the hypersymplectic setting, hence to the quaternionic case; hopefully this paper can also act as a motivation to study this problem.


\end{document}